\documentclass{emulateapj}

\begin{document}                  

\title{A {\it Spitzer}/IRS Spectrum of the 2008 Luminous Transient in
NGC~300: \\Connection to Proto-Planetary Nebulae}

\author{Jos{\'e}~L. Prieto\altaffilmark{1,3},
  Kris~Sellgren\altaffilmark{1}, Todd~A.~Thompson\altaffilmark{1,2,4},
  Christopher~S.~Kochanek\altaffilmark{1,2}}

\altaffiltext{1}{Dept.\ of Astronomy, The Ohio State University, 
140 W.\ 18th Ave., Columbus, OH 43210}

\altaffiltext{2}{Center for Cosmology and AstroParticle Physics, 
The Ohio State University, 191 W.\ Woodruff Ave., Columbus, OH 43210}

\altaffiltext{3}{Now Hubble and Carnegie-Princeton Fellow at Carnegie
  Observatories, 813 Santa Barbara St., Pasadena, CA 91101}

\altaffiltext{4}{Alfred P.~Sloan Fellow}

\shorttitle{A {\it Spitzer}/IRS Spectrum of the 2008 Luminous Transient in
NGC~300}

\begin{abstract}
We present a {\it Spitzer}/IRS low-resolution mid-infrared spectrum
($5-14$~$\mu$m) of the luminous transient discovered in the nearby
galaxy NGC~300 in May 2008. This transient had peak luminosity $M_{V}
\simeq -13$, showed an optical spectrum dominated by relatively narrow
Balmer and Ca~II lines in emission, and its progenitor was identified in
pre-explosion images as a dust-enshrouded $\sim 10$~M$_{\odot}$ star,
characteristics that make it a twin of SN~2008S. The {\it Spitzer}
spectrum, obtained three months after discovery, shows that the
transient is very luminous in the mid-IR. Furthermore, the spectrum
shows strong, broad emission features at $8$~$\mu$m and $12$~$\mu$m that
are observed in Galactic carbon-rich proto-planetary nebulae. Combining
these data with published optical and near-IR photometry obtained at the
same epoch, we find that the mid-IR excess traced by the {\it Spitzer}
spectrum accounts for $\sim 20\%$ of the total energy output. This
component can be well explained by emission from $\sim
3\times 10^{-4}$~M$_{\odot}$ of pre-existing progenitor dust at temperature
${\rm T} \sim 400$~K. The spectral energy distribution of the transient
also shows a near-IR excess that can be explained by emission from
newly-formed dust in the ejecta. Alternatively, both the near-IR and
mid-IR excesses can together be explained by a single pre-existing
geometrically thick dust shell. In light of the new observations
obtained with {\it Spitzer}, we revisit the analysis of the optical
spectra and kinematics, which were compared to the massive
yellow-hypergiant IRC+10420 in previous studies. We show that
proto-planetary nebulae share many properties with the NGC~300 transient
and SN~2008S. We conclude that even though the explosion of a massive
star (${\rm M} \gtrsim 10$~M$_{\odot}$) cannot be ruled out, an
explosive event on a massive (${\rm M}\sim 6-10$~M$_{\odot}$)
carbon-rich AGB/super-AGB or post-AGB star is consistent with all
observations of the transients and their progenitors presented thus far.
\end{abstract}



\section{Introduction}
\label{sec:intro}

An intriguing luminous optical transient was discovered in the nearby
galaxy NGC~300 (hereafter NGC~300-OT) by the amateur astronomer
B.~Monard on May 14.1, 2008 (Monard 2008). The transient was faint
compared to normal core-collapse supernovae, with an absolute magnitude
at discovery of $M_V\simeq -13$ (Bond et al.~2008). The optical spectrum
obtained by Bond et al.~(2008) close to discovery was dominated by
relatively narrow Hydrogen Balmer and Ca~II lines (infrared triplet and
forbidden doublet) in emission, as well as strong Ca~II H\&K in
absorption. Shortly after discovery, Berger \& Soderberg (2008) reported
strong upper limits on the optical luminosity of the progenitor star
obtained from deep archival {\it HST} data, which led them to suggest
that the progenitor was a low-mass main sequence star and the transient
was a stellar merger, similar to the red Galactic nova V838 Monocerotis
(e.g., Bond et al.~2003).

However, Prieto~(2008b) reported the discovery of a luminous
mid-infrared (mid-IR) progenitor to the transient in archival {\it
Spitzer} images. The progenitor was a luminous dust-enshrouded star,
whose spectral energy distribution was consistent with a black-body of
${\rm R}\simeq 300$~AU radiating at ${\rm T} \simeq 300$~K, with L$_{\rm
bol}\simeq 6\times 10^{4}$~L$_{\odot}$. This discovery showed that
NGC~300-OT was connected to an energetic explosion in a relatively low
mass $\sim 10$~M$_{\odot}$ star. The relatively low luminosity of the
transient compared to normal core-collapse supernovae, spectral
properties, and dust-enshrouded nature of the progenitor star, made
NGC~300-OT a ``twin'' of SN~2008S (Prieto et al.~2008a; Prieto~2008b),
which was discovered earlier in 2008 in the galaxy NGC~6946 (Arbour \&
Boles~2008; Stanishev et al.~2008; Chandra \& Soderberg~2008; Steele et
al.~2008; Yee et al.~2008; Wesson et al.~2008).

There have been a number of studies of NGC~300-OT and SN~2008S, and
the nature of these transients is still under debate (Prieto et
al.~2008a; Thompson et al.~2009; Smith et al.~2009; Bond et al.~2009;
Berger et al. 2009; Botticella et al.~2009; Wesson et al.~2009;
Gogarten et al.~2009; Patat et al.~2009; Kashi et al.~2009). Thompson
et al.~(2009) present and discuss various possible physical mechanisms
that can explain these transients and the likely range of
main-sequence masses of the dusty progenitor stars that are consistent
with the observations of NGC~300-OT and SN~2008S: (1) massive
white-dwarf birth (${\rm M_{ZAMS}} \approx 6-8$~M$_{\odot}$); (2)
electron-capture supernova (${\rm M_{ZAMS}} \approx 9$~M$_{\odot}$);
(3) intrinsically low-luminosity iron core-collapse supernova (${\rm
M_{ZAMS} }\approx 10-12$~M$_{\odot}$); and (4) massive star outburst
(${\rm M_{ZAMS}} \approx 10-15$~M$_{\odot}$). Recently, Kashi et
al.~(2009) proposed a mass transfer event from an extreme-AGB star to
a main-sequence star as the possible physical mechanism for
NGC~300-OT. Any of these potential scenarios suggests that these
transients are very important for our understanding of the evolution
of stars at the dividing line between ``high'' and ``low'' mass (i.e.,
$8-10$~M$_\odot$).

Here we report on a low-resolution mid-IR spectrum of NGC~300-OT
obtained with {\it Spitzer} on August 14, 2008, 93~days after the
discovery and 113~days after the first detection (Monard 2008). The
transient is luminous in the mid-IR spectral range and shows broad
emission features that we interpret as signs of carbon-rich dust,
similar to the spectra of carbon-rich proto-planetary nebulae in the
Galaxy. The paper is organized as follows. In \S\ref{sec:data}, we
discuss the observations and data reduction. In \S\ref{sec:analysis},
we present the analysis of the spectrum and spectral energy
distribution of the transient. In \S\ref{sec:discussion}, we discuss
the implications of our findings. Hereafter we adopt a distance of
$1.88$~Mpc to NGC~300 (Gieren et al.~2005).

\section{{\it Spitzer} Observations}
\label{sec:data}

We observed NGC~300-OT with the Short-Low (SL; $5.2-14\,\mu{\rm m}$,
$R=\Delta \lambda/\lambda = 60-120$) module of the Infrared
Spectrograph (IRS; Houck et al.~2004) on August 14.4, 2008 (UT). The
observations were obtained in staring mode as part of a {\it Spitzer}
Director's Discretionary Time (DDT) proposal (PID 487; AOR key
28139008). The ramp time was set to 60~sec in both SL orders (SL1:
$7.4-14\,\mu{\rm m}$, SL2: $5.2-7.5\,\mu{\rm m}$) and 10 cycles were
obtained, for a total exposure time of 1200~sec on source, including
the two nod positions, for a total of 20 images.


\begin{figure}[t]
\includegraphics[width=3.6in,clip=true]{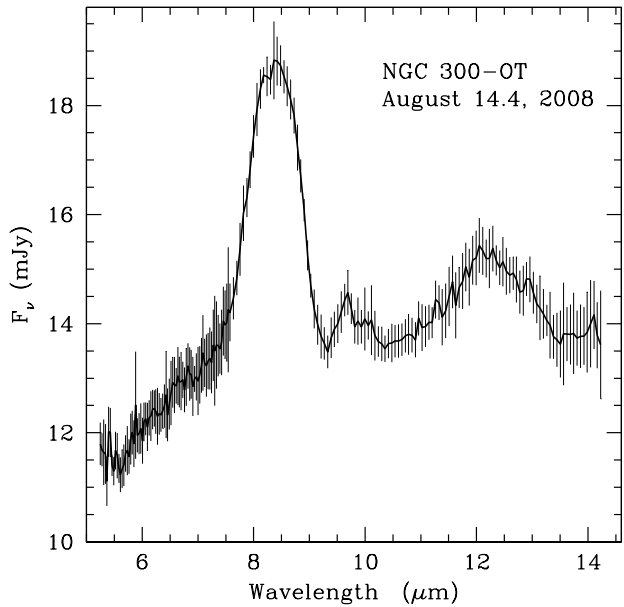}
\caption{Low-resolution {\it Spitzer}/IRS mid-IR spectrum of the
NGC300-OT obtained on 2008 August 14.4, $93$~days after the transient
was discovered. The vertical error bars are the RMS of each pixel
obtained after combining the 20 spectra.}
\label{fig:irs_spectrum}
\end{figure}



\begin{table}[!ht]
\begin{center}
\caption{Features in the {\it Spitzer}/IRS Spectrum of
NGC300-OT \label{table:lines}}
\begin{tabular}{lccccc}
\hline \hline
\\
\multicolumn{1}{c}{} &
\multicolumn{1}{c}{$\lambda_{c}$} &
\multicolumn{1}{c}{Intensity} &
\multicolumn{1}{c}{FWHM} \\
\multicolumn{1}{c}{} &
\multicolumn{1}{c}{($\mu$m)} &
\multicolumn{1}{c}{($10^{-13}\,\,{\rm erg\,\,cm^{-2}\,\,s^{-1}}$)} &
\multicolumn{1}{c}{($\mu$m)} \\
\\
\hline
\\
& $8.33\pm 0.01$   & $2.27 \pm 0.20$ & $0.94 \pm 0.03$ \\
& $9.71\pm 0.06$   & $0.12 \pm 0.11$ & $0.49 \pm 0.14$ \\
& $12.16\pm 0.07$  & $0.39 \pm 0.14$ & $1.47 \pm 0.16$ \\
\\
\hline
\hline
\end{tabular}
\end{center}
\end{table}


To reduce the data, we started from the basic calibrated data (BCD) from
the {\it Spitzer} Science Center pipeline (S18.1.0). We constructed a
high S/N background image for each order and nod by median combining all
the other images. We subtracted the background from the individual 2D
images. Rogue pixels in the background-subtracted images were cleaned
with IRSCLEAN (v1.9). We used the routines {\tt profile}, {\tt ridge},
{\tt extract} and {\tt tune} in the {\it Spitzer} IRS Custom Extractor
(SPICE) software package in order to extract flux-calibrated 1D
spectra. The spectra of each nod were median combined and the two orders
were merged together after applying a small multiplicative correction
factor of 3.5\% to the SL1 spectra.

Figure~\ref{fig:irs_spectrum} shows the final combined spectrum with
$\pm 1\sigma$ error bars on the fluxes estimated from the RMS in each
pixel. The mean signal-to-noise ratio of the final spectrum is $\simeq
35$ per pixel.

\section{Analysis}
\label{sec:analysis}

\subsection{Spectral Features}
\label{subsec:features}

The {\it Spitzer} spectrum of NGC~300-OT presented in
Figure~\ref{fig:irs_spectrum} shows two prominent broad emission
features at $\approx 8.3$~$\mu$m and $\approx 12.2$~$\mu$m. The
8~$\mu$m flux of the host galaxy at the background pointing positions,
measured within the IRS extraction aperture, is $< 1$~mJy. Since the
8~$\mu$m flux of the transient is $\sim 14$~mJy at the time of the IRS
spectrum, uncertainties in the subtraction of emission features from
polycyclic aromatic hydrocarbons (PAHs) in the host galaxy cannot
account for the spectral features we observe. There is also a
relatively narrow, but resolved, fainter feature at $\approx
9.7$~$\mu$m. The significance of this faint feature is quite uncertain
and depends sensitively on the spectra used to obtain the final
combined spectrum.

The main properties of the emission features (central wavelength,
FWHM, and integrated fluxes) present in the mid-IR spectrum are shown
in Table~\ref{table:lines}. They were obtained after fitting Gaussians
to the continuum-subtracted spectrum. The continuum was modeled using
a high-order (6th) polynomial fit over the wavelength regions:
$\lambda \leq 7.5$~$\mu$m, $9.2-9.3$~$\mu$m, $10.25-10.6$~$\mu$m, and
$\lambda \geq 13.4$~$\mu$m. We obtain consistent results if we use a
spline function to model the continuum.

\begin{figure}[t]
\includegraphics[width=3.6in,clip=true]{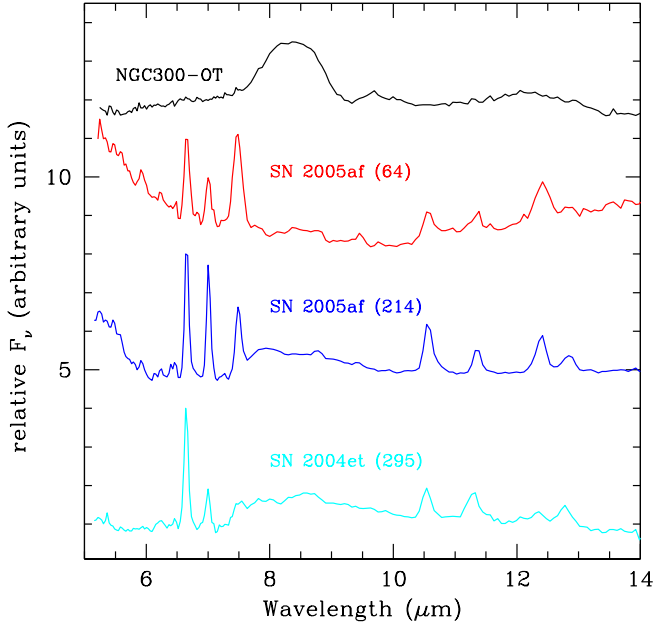}
\caption{Comparison of the {\it Spitzer}/IRS spectrum of NGC300-OT (top)
with mid-IR spectra of the type~IIP supernovae SN~2004et and SN~2005af.
In parenthesis are the days with respect to the explosion date for each
spectrum. We have subtracted a linear continuum fit in the wavelength
region shown to each spectrum and scaled the flux arbitrarily.}
\label{fig:comp1}
\end{figure}

In Figure~\ref{fig:comp1} we compare the {\it Spitzer} spectrum of
NGC~300-OT with mid-IR spectra of type~IIP supernovae. The spectra of
SN~2004et (Kotak et al.~2009) and SN~2005af (Kotak et al.~2006) were
obtained from the {\it Spitzer} archive (PID~237, 20256). Unlike the
late-time mid-IR spectra of normal type~IIP supernovae (e.g., SN~2004dj,
Kotak et al.~2005; SN~2005af, Kotak et al.~2006; SN~2004et, Kotak et
al.~2009) and SN~1987A (e.g., Roche et al.~1993; Wooden et al.~1993)
that are dominated by narrow fine-structure lines of stable Ni, Ar, Ne,
Co, and some molecular SiO in emission at $\sim 8-9$~$\mu$m, the mid-IR
spectrum of NGC~300-OT presents broad features that are most likely
dominated by emission from dust grains in the circumstellar
environment. The non-detection of fine-structure lines of Fe-peak
elements in emission suggests that the main source of dust heating at
this epoch is not the decay of radioactive $^{56}$Ni. This is consistent
with the low $^{56}$Ni production estimated by Botticella et al. (2009)
in the case of a supernova explosion.


\begin{figure}[t]
\includegraphics[width=3.6in,clip=true]{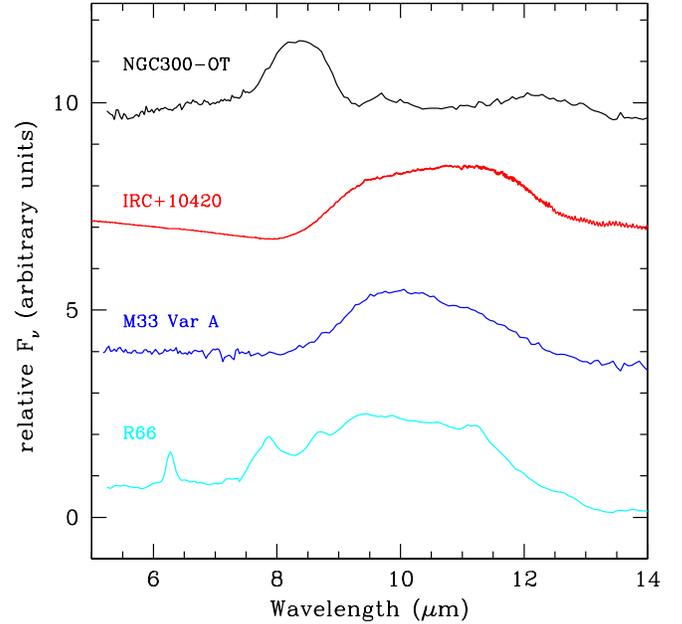}
\caption{Comparison of the {\it Spitzer}/IRS spectrum of NGC300-OT (top)
with mid-IR spectra of evolved massive stars that show circumstellar
dust emission. These include the yellow-hypergiants IRC+10420 and
M33~Var~A, and the B[e] supergiant R66. We have subtracted a linear
continuum fit in the wavelength region shown to each spectrum and scaled
the flux arbitrarily.}
\label{fig:comp2}
\end{figure}

In Figure~\ref{fig:comp2} we compare the {\it Spitzer} spectrum of
NGC~300-OT with mid-IR spectra of three evolved massive stars that
have circumstellar dust. The spectrum of the yellow-hypergiant
IRC+10420 (e.g., Humphreys et al.~2002) was obtained from the ISO
catalog of SWS spectra (Sloan et al.~2003). The spectra of the
yellow-hypergiant M33~Var~A (Humphreys et al.~2006) and the LMC B[e]
supergiant R66 (Kastner et al.~2006) are from the {\it Spitzer}
archive (PID~5, 3426). The spectra of IRC+10420, M33~Var~A and R66 are
dominated by the amorphous silicate emission feature at $9.7$~$\mu$m,
characteristic of oxygen-rich dust. The spectrum of R66 also contains
PAH emission features at 6.2, 7.7, 8.6 and 11.3~$\mu$m, indicating the
presence of carbon-rich dust as well. It is clear from
Figure~\ref{fig:comp2} that the mid-IR spectrum of NGC~300-OT does not
resemble the spectra of these evolved massive stars with circumstellar
dust, even though the optical spectrum of the transient is strikingly
similar to IRC+10420 (Bond et al.~2009; see Smith et al.~2009 for the
case of SN~2008S).


\begin{figure}[t]
\includegraphics[width=3.6in,clip=true]{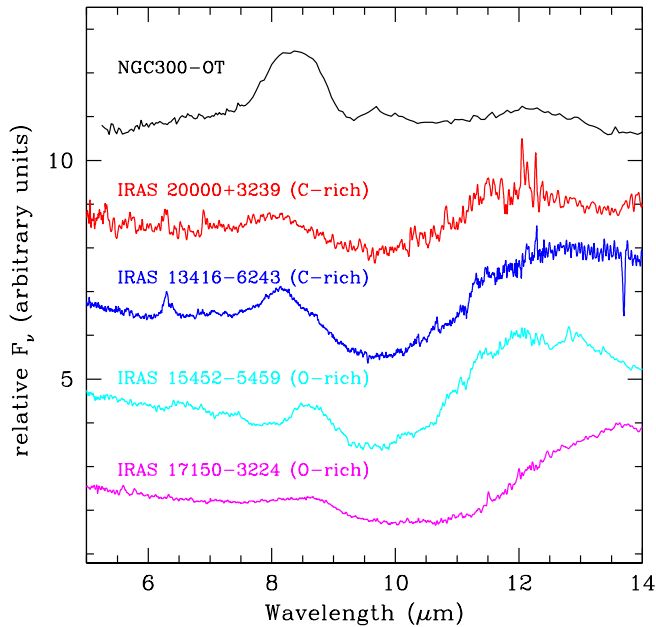}
\caption{Comparison of the {\it Spitzer}/IRS spectrum of NGC300-OT (top)
with mid-IR spectra of Galactic proto-planetary nebulae. The pPNe have
different dust chemistry: carbon-rich (IRAS~2000+3239 and
IRAS~13416-6243) and oxygen-rich (IRAS~1545-549 and
IRAS~17150-3224) dust. We have subtracted a linear continuum fit in
the wavelength region shown to each spectrum and scaled the flux
arbitrarily.}
\label{fig:comp3}
\end{figure}

In Figure~\ref{fig:comp3} we compare the spectrum of NGC~300-OT with
mid-IR spectra of Galactic proto-planetary nebulae (pPNe). Two of the
pPNe have carbon-rich dust (IRAS~20000+3239 and IRAS~13416-6243) and the
others have oxygen-rich dust (IRAS~15452-5459 and IRAS~17150-3224). The
spectra are all from the ISO/SWS catalog (Sloan et al.~2003). The
spectra of oxygen-rich pPNe contain SiO absorption at $7.9$~$\mu$m and a
strong silicate absorption feature at $9.7$~$\mu$m, which do not appear
to be present in the spectrum of NGC~300-OT. Also the central
wavelengths and FWHM of the two ``bumps'' at $\sim 8.5$~$\mu$m and
$\gtrsim 12$~$\mu$m are inconsistent with the features in the spectrum
of NGC~300-OT.

The spectrum of NGC~300-OT is most similar to the spectra of the
carbon-rich pPNe in Figure~\ref{fig:comp3}. They contain broad
emission features at $\sim 8$~$\mu$m and $\sim 12$~$\mu$m, which have
been associated with C-C and C-H bending and stretching modes
identified as the carriers of PAHs (e.g., Duley \& Williams~1981;
Peeters et al.~2002). Note, however, that the spectrum of NGC~300-OT
does not contain the 6.2~$\mu$m PAH feature that is clearly present in
the two carbon-rich pPNe. We can put a 3$\sigma$ upper limit on the
integrated flux of the 6.2~$\mu$m PAH feature of $I_{6.2} < 2.1\,({\rm
FWHM}/ 0.2\,\mu {\rm m})\times 10^{-14}$~erg~cm$^{-2}$~s$^{-1}$. This
gives a 3$\sigma$ limit on the flux ratio of $I_{6.2}/I_{8.3} < 0.09$
for an assumed ${\rm FWHM}= 0.2$~$\mu$m. The spectrum of
IRAS~20000+3239 also has the 6.9~$\mu$m PAH feature, which is not
present in the spectrum of NGC~300-OT. We can put a 3$\sigma$ upper
limit on the integrated flux of the 6.9~$\mu$m PAH feature of $I_{6.9}
< 1.9\,({\rm FWHM}/ 0.2\,\mu {\rm m})\times
10^{-14}$~erg~cm$^{-2}$~s$^{-1}$, which gives a limit on the flux
ratio $I_{6.9}/I_{8.3} < 0.08$ for an assumed ${\rm FWHM}=
0.2$~$\mu$m.

\subsection{Spectral Energy Distribution}
\label{subsec:SED}

\begin{figure}[!t]
\includegraphics[width=3.6in,clip=true]{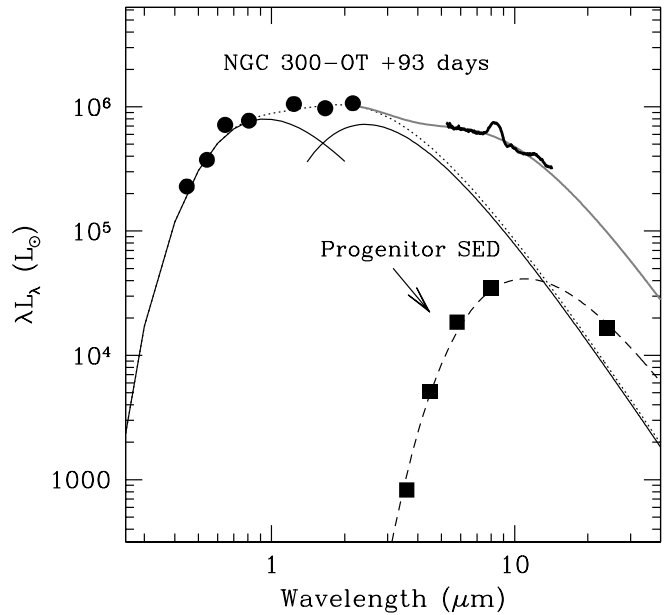}
\caption{SED of NGC~300-OT at 93~days after discovery. The {\it filled
circles} are the optical ($BVRI$) and near-IR ($JHK$) fluxes of
NGC~300-OT from Bond et al.~(2009). The {\it Spitzer}/IRS spectrum is
shown as the {\it thick black line}. The {\it dotted line} is a fit to
the optical+near-IR SED of the transient using the sum of two
black-bodies with temperatures T$_{1}\approx 3890$~K and T$_{2}\approx
1500$~K. The {\it grey line} is an extension of the previous black-body
fits to mid-IR by adding a cooler component with T$_{3} \approx
485$~K. We show the SED of the progenitor of NGC~300-OT for comparison
({\it filled squares}). The {\it dashed line} is a single black-body fit
with ${\rm T} \approx 335$~K to the progenitor SED.}
\label{fig:sed}
\end{figure}

We can construct the full spectral energy distribution (SED) of
NGC~300-OT at the epoch of the {\it Spitzer} spectrum using the
optical and near-IR photometry presented in Bond et
al.~(2009). Figure~\ref{fig:sed} shows the optical to mid-IR SED of
the transient 93~days after the discovery date. We have corrected all
the fluxes for a total extinction along the line-of-sight of
$E(B-V)=0.25$~mag, which is the mean of the extinction values reported
in Bond et al.~(2009). We assume $R_V=3.1$ and use the Schlegel et
al.~(1998) reddening law, with $A_{\lambda}\propto \lambda^{-1.6}$ in
the near-to-mid infrared range. The {\it filled circles} are the
optical ($BVRI$) and near-IR ($JHK$) fluxes from Bond et
al.~(2009). The thick line is the {\it Spitzer} mid-IR spectrum. For
comparison, we show the SED of the luminous dust-enshrouded progenitor
of NGC~300-OT ({\it filled squares}) obtained from pre-explosion {\it
Spitzer} IRAC ($3.6-8$~$\mu$m) and MIPS ($24$~$\mu$m) photometry (see
Table~\ref{table:phot_prog}; these data were used in Thompson et
al.~2009 and Bond et al.~2009). At the epoch of the {\it Spitzer}
observation the transient is $\sim 20$ times more luminous than the
progenitor at $8$~$\mu$m. The results of black-body fits to the
transient and progenitor SED using dust emissivity law $Q_{\lambda}
\propto \lambda^{-n}$ ($n=0,\,1$) are presented in
Table~\ref{table:sed}.


The evolution of the light curves of NGC~300-OT in different filters
presented by Bond et al.~(2009) shows that the transient becomes redder
in time, with the color evolving from $V-K \simeq 3.1$~mag at discovery
to $V-K\simeq 5.2$~mag at the time of the {\it Spitzer} spectrum. This
fast evolution in the $V-K$ color, while the $B-V$ color only changes
from $\simeq 0.8$~mag to $\simeq 1.1$~mag in the same time period,
suggests the presence of warm circumstellar dust formed in the explosion
or heated pre-existing dust. 

Botticella et al.~(2009) analyzed the SED of SN~2008S and showed that
the evolution in optical+near-IR fluxes could be explained with a
single ``hot'' black-body until $\sim 120$~days after explosion, but
they needed a second ``warm'' black-body component at later times. They
concluded that the near-IR flux excess of SN~2008S at $\gtrsim 120$~days
after explosion was possibly due to newly-formed dust in the ejecta or
shock-heated dust in the circumstellar environment. 


As in the case of SN~2008S, we find that the $BVRIJHK$ fluxes of
NGC~300-OT at 93~days after discovery (113~days after the first
detection\footnote{We assume this is a good estimate of the time after
explosion. This seems like a reasonable assumption given that the
light curve of the transient is rising fast at that time (see Figure~1
in Bond et al.~2009).} ) can be well-fit by the sum of two
black-bodies ($\chi^2_{\nu} = 1.8$), a hot component with T$_{1}
=3893$~K, R$_{1} = 10.7$~AU and a warm component with T$_{2}=1511$~K,
R$_{2} = 67.3$~AU (see the dotted line in Figure~\ref{fig:sed}). The
total luminosity of these components is $2.1\times
10^{6}$~L$_{\odot}$. A modified black-body with dust emissivity
$Q_{\lambda} \propto \lambda^{-1}$ for the colder component gives
T$_{1}=3870$~K, T$_{2}=1256$~K, and comparable total
luminosity. However, the mid-IR SED of NGC~300-OT traced by the {\it
Spitzer} spectrum is a factor of $\sim 2-9$ brighter between
$5-14$~$\mu$m than the extrapolated sum of the two black-bodies. We
add a third black-body component with lower temperature, while keeping
the fit to the optical and near-IR fluxes fixed, to account for the
mid-IR excess. We find a good fit ($\chi^2_{\nu} = 1.5$) to the
continuum of the {\it Spitzer} spectrum (defined in
\S\ref{subsec:features}) with T$_{3} = 485$~K and R$_{3} =
515$~AU. The luminosity of this component is $6.0\times
10^{5}$~L$_{\odot}$, which is $29\%$ of the optical+near-IR luminosity
and $22\%$ of the total integrated luminosity of the transient. A
modified black-body with dust emissivity $Q_{\lambda} \propto
\lambda^{-1}$ gives T$_{3}=394$~K and total luminosity $6.4\times
10^{5}$~L$_{\odot}$.
 
The two black-body components that can reproduce the near-IR and
mid-IR excesses in the SED of NGC~300-OT are likely due to emission
from circumstellar dust. The hotter component (T$_2=1511$~K) can be
reasonably explained with newly formed dust in the ejecta, as proposed
by Botticella et al.~(2009) for SN~2008S. The velocity inferred from
the black-body radius R$_2 = 67$~AU and the time after the first
detection of the transient is $\sim 1000$~km~s$^{-1}$. This velocity
is in the range of velocities measured from emission lines in the
optical spectra of NGC~300-OT of $\sim 100-1000$~km~s$^{-1}$ (Bond et
al.~2009; Berger et al.~2009).

Another way to explain this warm dust component would be emission from
pre-existing progenitor dust, although this seems less likely because of
dust destruction from the initial outburst light (e.g.,
Dwek~1983). Assuming a luminosity at maximum light of L$_{\rm max}
\approx 5\times 10^{6}$~L$_{\odot}$ (luminosity at discovery) and a dust
sublimation temperature of ${\rm T_{sub}} \simeq 1500$~K, we obtain a
radius for the dust-free cavity of $\sim 80-250$~AU depending on the
assumed dust emisivity law, $Q_{\lambda} \propto \lambda^{0} -
\lambda^{-1}$. This radius is a factor of $\sim 1.2-3.7$ times larger
than the black-body scale R$_2$ of the $\sim 1500$~K temperature dust,
which suggest the near-IR emitting dust may have been formed in the
ejecta.


\begin{table}[!t]
\begin{center}
\caption{Spectral Energy Distribution of the Progenitor of NGC~300-OT
\label{table:phot_prog}}
\begin{tabular}{lccccc}
\hline \hline
\\
\multicolumn{1}{c}{} &
\multicolumn{1}{c}{$\lambda$} &
\multicolumn{1}{c}{} &
\multicolumn{1}{c}{$\lambda F_{\lambda}$} &
\multicolumn{1}{c}{Telescope/Instrument} \\ 
\multicolumn{1}{c}{} &
\multicolumn{1}{c}{} &
\multicolumn{1}{c}{} &
\multicolumn{1}{c}{($10^{-14}\,\,{\rm erg\,\,cm^{-2}\,\,s^{-1}}$)} &
\multicolumn{1}{c}{} \\
\\
\hline
\\
& 3.6~$\mu$m   &  & $0.75  \pm 0.15$  & {\it Spitzer}/IRAC \\ 	
& 4.5~$\mu$m   &  & $4.67  \pm 0.43$  & {\it Spitzer}/IRAC \\
& 5.8~$\mu$m   &  & $16.83 \pm 1.10$  & {\it Spitzer}/IRAC \\
& 8.0~$\mu$m   &  & $31.49 \pm 1.67$  & {\it Spitzer}/IRAC \\
& 24~$\mu$m    &  & $15.03 \pm 1.37$  & {\it Spitzer}/MIPS \\	
\\
\hline
\hline
\end{tabular}
\end{center}
\end{table}


We can estimate the total dust mass needed to account for the luminosity
of the warm black-body component using equation~(4) in Dwek et
al.~(1983) and assuming an average carbon grain density of $\rho_{\rm
dust} \simeq 2.24$~g~cm$^{-3}$,

\begin{equation}
{\rm M_{d}} \approx 2 \times 10^{-6} \left (\frac{{\rm
T_{d}}}{1000\,{\rm K}} \right )^{-5} \left ( \frac{{\rm L_{d}}}{10^{6}
{\rm L}_{\odot}} \right ) \,{\rm M}_{\odot}
\label{eq:dustmass}
\end{equation}

\noindent where T$_d$ is the dust temperature in Kelvin, and L$_d$ is
the dust luminosity in L$_{\odot}$. This equation assumes that the
dust grains have an absorption/emission efficiency of $Q_{\lambda}
\propto \lambda^{-1}$. Using T$_2$ and L$_2$ for $n=1$ in
Table~\ref{table:sed}, we obtain ${\rm M_{d}} \approx
10^{-6}$~M$_{\odot}$ for the mass of newly formed dust at 93~days
after discovery. We can compare this dust mass to that in
SN~2008S. Under the same assumptions about dust properties considered
in equation~(\ref{eq:dustmass}), this dust mass is $\sim 7$ times
larger than was needed to explain the near-IR excess in SN~2008S $\sim
120$~days after explosion (using the data in Table~8 of Botticella et
al.~2009).

The mid-IR excess revealed by the {\it Spitzer} spectrum of NGC~300-OT
cannot be explained by newly formed dust. The constant expansion
velocity needed to reach a black-body radius of R$_3=515$~AU at this
epoch is $v\sim 8000$~km~s$^{-1}$, far larger than the velocities
observed in the optical spectra of the transient. This component must
be emission from pre-existing dust from the progenitor. Dust grains
that were not destroyed by the initial outburst will absorb the
outburst light, warm up, and re-radiate at mid-IR wavelengths. Using
equation~(\ref{eq:dustmass}) we estimate that a dust mass of ${\rm
M_{d}} \approx 3\times 10^{-4}$~M$_{\odot}$ and dust optical depth of
$\tau_V \approx 0.4$ is needed to account for the mid-IR excess,
assuming dust emissivity with $n=1$ in Table~\ref{table:sed}. For a
gas-to-dust mass ratio of 200 that is consistent with observations of
evolved stars with carbon-rich dust in the LMC (e.g., Matsuura et
al.~2009), we find a gas mass of $\approx 0.06$~M$_{\odot}$. A similar
mass of dust of $\sim 10^{-4}$~M$_{\odot}$ is needed to explain the
SED of the progenitor of NGC~300-OT, suggesting that a substantial
fraction of the dust in the progenitor wind survives the
explosion. Interestingly, a similar mass of dust is also needed to
explain the mid-IR excess observed in SN~2008S $\sim 17$~days after
explosion (Botticella et al.~2009).


\begin{table}[!t]
\begin{center}
\caption{Black-body Fits to the Transient \\ and Progenitor SEDs
\label{table:sed}}
\begin{tabular}{lcrcr}
\hline \hline
Parameter & & Value\tablenotemark{a} ($n=0$)  && Value\tablenotemark{a} ($n=1$) \\
\hline
\multicolumn{5}{c}{NGC~300-OT $BVRIJHK$ fluxes}
\\
\hline
T$_{1}$~(K) & & 3893 & & 3870 \\ 	
R$_{1}$~(AU) & &  10.7 & & $\ldots$ \\
L$_{1}$~(L$_{\odot}$) & &  $1.1\times 10^{6}$ & & $1.1\times 10^{6}$ \\	
\hline
T$_{2}$~(K) & &  1511 & &  1256 \\
R$_{2}$~(AU) & &  67.3 & & $\ldots$ \\
L$_{2}$~(L$_{\odot}$) & &  $9.8\times 10^{5}$ & & $8.2\times 10^{5}$ \\
\hline 
\multicolumn{5}{c}{NGC~300-OT adding fit to {\it Spitzer} data}
\\
\hline
T$_{3}$~(K) & &   485 & & 394 \\ 	
R$_{3}$~(AU) & &   515 & & $\ldots$ \\
L$_{3}$~(L$_{\odot}$) & &   $6.0\times 10^{5}$ & & $6.4\times 10^{5}$ \\
\hline
\multicolumn{5}{c}{Progenitor SED}
\\
\hline
T~(K) & &   335 & &  281 \\ 	
R~(AU) & &   332 & & $\ldots$ \\
L~(L$_{\odot}$) & &   $5.7\times 10^{4}$ & &  $5.4\times 10^{4}$ \\
\hline \hline
\end{tabular}
\tablenotetext{a}{Black-body fits with dust emissivity law $Q_{\lambda} \propto \lambda^{-n}$.}
\end{center}
\end{table}


We performed radiation transport calculations using DUSTY (Ivezic \&
Elitzur~1997) to check if the SED of NGC~300-OT can also be explained by
radiation through a spherical shell of dust. We found a reasonable fit
to the optical to mid-IR SED using a central black-body with
T$=4000-5000$~K illuminating a dust shell with density profile $\rho(r)
\propto r^{-1}$, inner and outer radius of the shell R$_{\rm in} \simeq
100$~AU and R$_{\rm out}\simeq 10000$~AU, temperature of the dust at the
inner radius $T=1500$~K, and dust optical depth $\tau_V \simeq 1.2$. We
note that this solution is not unique since we also obtain a reasonable
fit to the SED with $\rho(r)= {\rm constant}$, R$_{\rm in} \simeq
100$~AU, and ${\rm R_{out}/R_{in} }\sim 10$, with all the other
parameters being equal. Wesson et al.~(2009) analyzed the optical to
mid-IR SED of the progenitor of SN~2008S and the transient $\sim
17$~days and $\sim 180$~days after explosion using a radiative transfer
code. They find that the SED of SN~2008S can be explained with a central
source illuminating a spherical dust shell with density profile $\rho
\propto r^{-2}$, inner radius R$_{\rm in} \simeq 1250$~AU, and $\tau_{V}
\simeq 0.8$, where $\gtrsim 98\%$ of the pre-existing progenitor dust
(M$_{\rm dust}\simeq 1.2\times 10^{-5} - 3.5\times 10^{-3}\, {\rm
M}_{\odot}$, depending on R$_{\rm out}$) survives the explosion. This is
qualitatively similar to our results for NGC~300-OT.

\section{Discussion \& Conclusions}
\label{sec:discussion}

We have presented a low-resolution mid-IR spectrum of NGC~300-OT
obtained with {\it Spitzer} on August 14.4, 2008, 93~days after the
discovery of the transient. We now give our discussion and
interpretation of the results.

\subsection{Mid-IR Spectrum and SED of NGC~300-OT}

The mid-IR spectrum of NGC~300-OT shows broad emission features at
$8.3$~$\mu$m and $12.2$~$\mu$m that are similar to the broad features
seen in the spectra of carbon-rich pPNe in the Galaxy (e.g., Kwok 1993;
Kwok et al.~2001), called ``Class C'' PAH sources by Peeters et
al.~(2002). Joblin et al.~(2008) derive profiles for the broad 8~$\mu$m
and 12~$\mu$m features from the spectrum of the archetypal ``Class C''
PAH source, the pPN IRAS~13416-6243 (see Figure~4). These broad features
are attributed to hydrocarbons with a predominantly aliphatic nature,
which undergo photochemical processing in proto-planetary nebula to
transform into the more aromatic material observed in carbon-rich
planetary nebulae (e.g., Kwok et al.~2001; Sloan et al.~2007). Joblin et
al.~(2008) show that these broad features are also observed in young
planetary nebulae, and are distinct from the spectral features of
neutral PAHs, ionized PAHs, and very small grains.

It is interesting to note that the position of the center of the PAH
complex at $7-8$~$\mu$m, observed in many astrophysical environments
including the ISM and evolved stars (e.g., Tielens 2008 and references
therein), has been shown to correlate with the effective temperature of
the host star in Herbig A/Be stars, planetary-nebulae, and pPNe (e.g.,
Sloan et al.~2007; Keller et al.~2008; Boersma et~al. 2008). The
correlation goes in the sense that stars with lower effective
temperatures (i.e., weaker UV-optical radiation field) show the central
peak of this complex at longer wavelengths. The central wavelength at
$8.3$~$\mu$m detected in the {\it Spitzer} spectrum of NGC~300-OT would
imply an effective temperature of $\sim 4000$~K (see Fig.~8 in Keller et
al.~2008), which is consistent with the temperature of the hot
black-body ($T\simeq 3900$~K) derived from the optical SED of the
transient. This provides indirect and independent support for our
interpretation of the mid-IR spectrum, and also evidence that UV
processing has not yet converted predominantly aliphatic hydrocarbons
into PAHs in NGC~300-OT.

A noteworthy difference between the {\it Spitzer} spectrum of NGC~300-OT
and the spectra of carbon-rich pPNe is the non-detection of the
6.2~$\mu$m PAH emission feature to fairly deep limits. In the
carbon-rich pPN IRAS~13416-6243, for example, the ratio of the
integrated fluxes of the 6.2~$\mu$m and 8~$\mu$m features is
$I_{6.2}/I_{8} \simeq 0.13$, which is a factor of $1.3$ higher than the
3$\sigma$ limit for NGC~300-OT. The 6.2~$\mu$m emission feature is
thought to be produced by the C-C stretching mode in ionized PAHs. The
astronomical and laboratory spectra of PAHs and PAH-like molecules show
such a wide variety that the absence of the 6.2~$\mu$m feature may be
explained by differences in shape, ionization state, impurities, and
size of the molecules (e.g., Pathak \& Rastogi~2008; Bauschlicher et
al.~2009). On the other hand, the non-detection of the 6.9~$\mu$m PAH
emission feature in the spectrum of NGC~300-OT seems to be consistent
with the integrated flux of this feature measured in some Galactic
carbon-rich pPNe like IRAS~20000+3239. This emission feature is thought
to be associated with aliphatic material (C-H bending mode) and is
detected only in a fraction of pPNe (e.g., Kwok et al.~1999).

The mid-IR excess traced by the {\it Spitzer} spectrum can be well
explained by the presence of warm circumstellar dust (${\rm T}\sim
400$~K) with mass M$_d \sim 3\times 10^{-4}$~M$_{\odot}$. This dust
was most likely part of the dusty progenitor wind, pre-existing the
luminous explosion that produced the optical transient. The SED of
NGC~300-OT at the epoch of the {\it Spitzer} spectrum also shows a
near-IR excess, which can be explained with a small mass ($\sim
10^{-6}$~M$_{\odot}$) of warm circumstellar dust (${\rm T}\sim
1300$~K) formed in the ejecta.

Alternatively, the SED of NGC~300-OT can be reasonably well explained
with a ${\rm T}=4000-5000$~K black-body illuminating a spherical shell
of pre-existing progenitor dust that extends from $\sim 100$~AU to $\sim
10000$~AU, where the inner radius of the dust shell marks the
destruction of dust by the initial outburst light. The presence of a
substantial mass of pre-existing dust from the progenitor wind in the
overall SED of NGC~300-OT was also characteristic of SN~2008S (Wesson et
al.~2009) and indicates that most of the dust survived the explosion.

\subsection{NGC~300-OT and SN~2008S: Connection to Proto-Planetary Nebulae}

The similarity of the mid-IR spectrum of NGC~300-OT with carbon-rich
pPNe is very striking and may shed new light on the nature of this
transient, SN~2008S, and the other optical transients (e.g., SN~1999bw,
M85-OT) that show similar characteristics and appear to be part of the
same class (Prieto et al.~2008c; Thompson et al.~2009). The optical
spectra of NGC~300-OT and SN~2008S transients were compared with the
spectrum of the massive Galactic yellow-hypergiant star IRC+10420 (Smith
et al.~2009; Bond et al.~2009; Berger et al.~2009), which shows an
F-type supergiant spectrum with Balmer lines in emission, as well as
strong Ca~II triplet, [Ca~II] doublet, and [Fe~II] lines in
emission. Given the similarity of the mid-IR spectrum of NGC~300-OT with
pPNe, we have searched in the literature for their optical spectra. We
found several examples of pPNe with optical spectra that are remarkably
similar to NGC~300-OT and SN~2008S. In the sample of echelle long-slit
spectra of evolved stars of S{\'a}nchez~Contreras et al.~(2008), the
Galactic pPNe IRAS~17516-2525 (O-B spectral type), M1-92 (B2-F5),
Hen~3-1475 (Be), IRAS~22036+5306 (F4-7), and IRAS~08005-2356 (F4~Ie)
show Balmer lines and also strong Ca~II triplet and [Ca~II] doublet in
emission. Other examples can be found in the atlas of optical spectra of
post-AGB stars presented in Su{\'a}rez et al.~(2006). The presence of
forbidden Ca~II in emission in the spectra of NGC~300-OT and SN~2008S,
pPNe, and IRC+10420, which is rarely present in stellar spectra, means
that calcium is not depleted onto dust grains, most likely due to the
destruction of grains by relatively fast shocks (e.g., Hartigan et
al.~1987).

Another similarity between NGC~300-OT and pPNe is revealed by the
kinematics and the detection of double-peaked Balmer and Ca~II triplet
lines in the spectrum of NGC~300-OT. Bond et al.~(2009) interpreted
these double features as the presence of a bipolar outflow with an
expansion velocity of $\approx 75$~km~s$^{-1}$, and possibly faster
components moving at $\sim 200$~km~s$^{-1}$. Berger et al.~(2009)
discussed evidence for even faster velocity components (including
inflow) going up to $\sim 1000$~km~s$^{-1}$. Aspherical winds or outflow
moving at velocities of a ${\rm few}\times 100$~km~s$^{-1}$ up to $\sim
1000$~km~s$^{-1}$ are observed in pPNe (e.g., Balick \&
Frank~2002). Multiple studies using high-resolution imaging of Galactic
pPNe with {\it HST} have shown a variety of complex morphologies, with
bipolar, multipolar and point-symmetric structures (e.g., Sahai et
al.~1999). In particular the pPN Hen~3-1475 (also classified as a young
PN in some studies), which is in the spectroscopic sample of
S{\'a}nchez~Contreras et al.~(2008) and whose spectrum shares many
features with the spectra of NGC~300-OT, has a bipolar morphology and
velocity components up to $\sim 1200$~km~s$^{-1}$ (e.g., Riera et
al.~1995, 2003). Given the inferred luminosity of $\sim
10^{4}$~L$_{\odot}$, chemistry, and kinematics, Riera et al.~(1995,
2003) proposed that Hen~3-1475 was a relatively high-mass star ($\sim
3-5$~M$_{\odot}$) in the post-AGB phase of evolution. Another
interesting example is the Red Rectangle, an extensively studied
intermediate-mass pPNe with carbon-rich dust chemistry and a fast ($\sim
560$~km~s$^{-1}$) bipolar outflow traced by H$\alpha$ in emission (e.g.,
Witt et al.~2009).

Several studies have proposed that SN~2008S and NGC~300-OT were the
result of an energetic eruption in a dust-enshrouded $10-20$~M$_{\odot}$
star, where the star survives the eruption. Smith et al.~(2009)
discussed a super-Eddington wind as the physical mechanism that produced
SN~2008S, similar to the super-outbursts of massive LBVs (e.g., van
Marle et al.~2008). Berger et al.~(2009) presented possible
observational evidence for this model from the complex kinematics that
they inferred from their high-resolution spectra of NGC~300-OT. Bond et
al.~(2009) did not require that the progenitor of the transient was
LBV-like, but rather an OH/IR star (e.g., Wood et al.~1992) that was
evolving to warmer temperatures (in a blue-loop) at the time of the
eruption. These studies relied heavily on comparing the optical spectra
of the transients with the spectrum of the massive yellow-hypergiant
IRC+10420 (e.g., Humphreys et al.~2002; Davies et al.~2007). However, as
discussed here, there are examples of pPNe in the Galaxy that share very
similar optical spectroscopic characteristics with NGC~300-OT and
SN~2008S. In fact, the complex model of inflow-outflow put forth in
Berger et al.~(2009) to explain the spectra of NGC~300-OT has been
discussed in the context of fast winds of AGB and post-AGB stars in
binaries (Soker 2008).  

Finally, in a mid-IR study of massive stars in the LMC, Bonanos et
al.~(2009) argued that B[e] supergiants (e.g., R66 in
Figure~\ref{fig:comp2}) may share a common origin with NGC~300-OT and
SN~2008S. Supergiant B[e] stars in the LMC are very rare (only $\sim 10$
discovered), have luminosities L$_{\rm bol} \gtrsim 10^{4}$~L$_{\odot}$,
and dusty circumstellar envelopes, properties that are broadly
consistent with the properties of the progenitors of NGC~300-OT and
SN~2008S. However, the circumstellar dust around B[e] supergiants in the
LMC is significantly hotter ($\gtrsim 800$~K) than the dust around the
progenitors, probably because they are oxygen-rich.

In summary, we have shown that NGC~300-OT and SN~2008S have several
properties (mid-IR spectrum, optical spectra, kinematics, and dusty
circumstellar medium) that are characteristic of pPNe in the Galaxy;
they are not unique to massive stars like IRC+10420.

\subsection{The Progenitors of NGC~300-OT and SN~2008S: Massive Carbon-rich AGB/post-AGB stars ?}

The progenitors of NGC~300-OT and SN~2008S were luminous ($\sim
4-6\times10^{4}$~L$_{\odot}$) dust-enshrouded stars with warm (T$\sim
300-450$~K) circumstellar dust, found at the red extremum of the AGB
sequence in a mid-IR color-magnitude diagram (Thompson et
al.~2009). They are part of the extreme-AGB (EAGB) sequence, which has
been identified as a continuation of the AGB to redder mid-IR colors
in resolved stellar population studies of nearby galaxies using {\it
  Spitzer} (e.g., LMC, Blum et al.~2006; M33, Thompson et
al.~2009). Their location in the mid-IR color-magnitude diagram
indicates extreme mass-loss and relatively cool circumstellar dust
(e.g., Srinivasan et al.~2009). Interestingly, Matsuura et al.~(2009)
find that most EAGB stars in the LMC sample for which they have
obtained mid-IR spectra have carbon-rich dust, which is consistent
with the evidence presented here for carbon-rich dust in
NGC~300-OT. Even though the number of carbon-rich AGBs in the LMC
declines as a function of luminosity with respect to oxygen-rich AGBs,
interpreted as evidence of Hot-Bottom-Burning which converts carbon
into nitrogen and oxygen, there are carbon stars with luminosities
approaching those of the progenitors of NGC~300-OT and SN~2008S (e.g.,
van Loon et al.~1998; Frost et al.~1998). One example is
IRAS~05278-6942, a carbon-rich AGB star in the LMC that has L$_{\rm
  bol} \sim 4\times 10^{4}$~L$_{\odot}$ and $\dot{\rm M} \sim 3\times
10^{-5}$~M$_{\odot}$~yr$^{-1}$ (Groenewegen et al.~2007). Indeed,
Kastner et al.~(2008) in their study of the most luminous 8~$\mu$m
sources in the LMC, point out that ``more high-L$_{\rm bol}$ carbon
stars may lurk among the very red, unclassified objects'' in their
sample.

The high-mass counterparts of AGB stars with M$_{\rm ZAMS}\simeq
8-10$~M$_{\odot}$, so-called super-AGB stars, have been proposed as good
candidates for the progenitors of NGC~300-OT and SN~2008S (see Thompson
et al.~2009 and references therein; Botticella et al.~2009). These stars
end up with an O-Ne core and, depending on the competing effects of
core-growth after carbon ignition and strong mass-loss, they can explode
as electron-capture supernovae in a narrow and uncertain mass range
around $\sim 9$~M$_{\odot}$ or end-up as O-Ne white dwarfs at lower
masses (e.g., Nomoto~1984; Poelarends et al.~2008). The luminosities of
SAGB stars in theoretical models can reach $\sim 10^{5}$~L$_{\odot}$ at
the end of their evolution (e.g., Siess~2007), comparable to the
luminosities of the progenitors of NGC~300-OT and SN~2008S. These models
also predict that the photospheric abundances of SAGB stars should be
oxygen-rich (${\rm C/O} < 1$) at the end of their evolution, through a
combination of Hot-Bottom-Burning and the occurrence of the third
dredge-up. However, the modeling of these processes in the AGB and SAGB
evolution is very uncertain and depends on several important factors
like metallicity, the treatment of convection, mass-loss, and the input
opacities (e.g., Marigo~2008). In fact, there are theoretical studies
that have discussed the possibility of carbon-rich photospheres in
massive AGBs (e.g., Nomoto~1987; Marigo~2007).

An important difference between the progenitors of the luminous
transients and carbon-rich AGB and EAGB stars is that they did not show
variability in the mid-IR within $3-4$~years of explosion (Prieto et
al.~2008b; Thompson et al.~2009), whereas most AGB and EAGB stars are
highly variable (e.g., Gronewegen et al.~2007; Vijh et al.~2009). Since
variability in AGB stars is explained by pulsations that drive the
mass-loss (thermal pulses), the lack of variability in the progenitors
may indicate that they were at the very tip of the AGB or SAGB phase
before the explosion, perhaps past a {\it super-wind} phase. If this is
the case, the progenitor could be classified as a pPN (i.e., it was in
the post-AGB phase at the time of the explosion).

\subsection{Progenitors and Transients: Concluding Remarks}

The physical mechanism that produced the energetic explosions ($\sim
2-6\times 10^{47}$~erg in optical to near-IR light) of NGC~300-OT and
SN~2008S is still unknown. Although the observations presented here do
not directly shed light on the mechanism that produced the transients,
we have shown that all the observations of the transients and their
progenitors presented thus far are consistent with the explosion of a
massive (M$_{\rm ZAMS}\sim 6-10$~M$_{\odot}$), carbon-rich AGB,
super-AGB or post-AGB star, either single or in a binary. An in-depth
discussion of some of the mechanisms that could explain the transients
can be found in Thompson et al.~(2009). Here we briefly comment on the
ones that involve a massive AGB or post-AGB star: white dwarf formation
and an electron-capture supernova.
 
In the case of an energetic eruption where the progenitor survives the
explosion, the transients could mark the birth of massive white dwarfs
(Thompson et al.~2009). Observations of mass-losing AGB stars show
spherically symmetric envelopes, while their descendants
(proto-planetary and planetary nebulae) have highly asymmetric and
complex morphologies and kinematics. This has been a long standing
mystery in stellar evolution for which several mechanisms have been
proposed, with magnetic fields, rotation and binaries suggested as
primary suspects for breaking the symmetry (e.g., Balick \& Frank
2002; de Marco~2009; Sahai~2009 and references therein). In a recent
study, Dennis et al.~(2008) argue that pPNe outflows may be driven by
an explosive MHD launch mechanism similar to the ones discussed in the
context of supernovae and gamma-ray bursts. This model seems appealing
when applied to NGC~300-OT and SN~2008S -- perhaps we are witnessing
the launch of the jet in a massive AGB which is shaping a pPN. In this
scenario we expect that the pPN now in formation will become a PN when
the central white dwarf left behind ionizes the surrounding gas. The
timescale for this is very uncertain, but for a $\sim 8$~M$_{\odot}$
star it can be of the order of $\sim 100$~yr (e.g., Stanghellini \&
Renzini~2000). An interesting prediction of an asymmetric outflow that
can be tested with new observations is the detection of strongly
polarized light from the transient, which has recently been reported
by Patat et al.~(2009).

An electron-capture supernova in a massive AGB star has been suggested
as a possible mechanism for NGC~300-OT and SN~2008S (e.g., Prieto et
al. 2008b; Thompson et al.~2009; Botticella et al.~2009). Two of the
main predictions of this scenario that can be tested with late-time
observations are the disappearance of the progenitor star years after
the explosion and the detection of radioactive $^{56}$Ni decay
synthesized in the explosion. Botticella et al.~(2009) presented
detailed photometric and spectroscopic observations of SN~2008S. Their
main argument in favor of a supernova explosion as the origin of the
transient was presented in the late-time light curve. They found that
the pseudo-bolometric light curve at $t \gtrsim 140$~days had a decay
slope consistent with radioactive decay of $^{56}{\rm Co} \rightarrow$
$^{56}{\rm Fe}$ and inferred the production of $\sim
10^{-3}$~M$_{\odot}$ of $^{56}$Ni in the explosion. However, Smith et
al. (2009) noted that the late time light curve of SN~2008S was slower
than expected from $^{56}{\rm Co}$ decay and argued against a supernova
interpretation. While a late-time decay slope slower than $^{56}{\rm
Co}$ is a possibility in SN~2008S, it should be pointed out that slow
late-time light curve slopes (compared to $^{56}{\rm Co}$ decay) have
also been observed in some subluminous type~IIP supernovae, including
the very well-studied SN~2005cs (Pastorello et al.~2009). Therefore, we
think that a supernova explosion origin cannot be excluded from this
result alone. The late-time light curve of NGC~300-OT should give very
important clues about the possible supernova origin.


\acknowledgments

We thank M.~Barlow, J.~Beacom, A.~Bonanos, H.~Bond, R.~Humphreys,
N.~Smith, S.~Srinivasan K.~Stanek, and L.~Watson for many helpful
discussions and suggestions about these intriguing transients and
their progenitors. We thank the {\it Spitzer Science Center} Director,
T.~Soifer, for granting us Director's Discretionary Time to obtain the
IRS spectrum, and S.~Laine from the {\it SSC} for helping prepare the
observations. This work is based on observations made with the {\it
  Spitzer Space Telescope}, which is operated by the Jet Propulsion
Laboratory, California Institute of Technology under a contract with
NASA. JLP is supported by NSF grant AST-0707982. KS thanks NASA for
providing support for this work through an award issued by
JPL/Caltech. TAT is supported in part by an Alfred P.~Sloan
Fellowship. J.L.P. acknowledges support from NASA through Hubble
Fellowship grant HF-51260.01-A awarded by the STScI, which is operated
by AURA, Inc. for NASA, under contract NAS 5-26555. This research is
supported in part by NSF grant AST-0908816.


\begin{thebibliography}{99}
\frenchspacing

\bibitem[Arbour \& Boles (2008)]{2008CBET.1234....1A} Arbour, R., \&
Boles, T. 2008, CBET, 1234, 1

\bibitem[Balick \& Frank(2002)]{2002ARA&A..40..439B} Balick, B., \&
Frank, A.\ 2002, \araa, 40, 439

\bibitem[Bauschlicher et al.(2009)]{2009ApJ...697..311B} Bauschlicher,
C.~W., Peeters, E., \& Allamandola, L.~J.\ 2009, \apj, 697, 311

\bibitem[Berger \& Soderberg(2008)]{2008ATel.1544....1B} Berger, E., \&
Soderberg, A.\ 2008, The Astronomer's Telegram, 1544, 1

\bibitem[Berger et al.(2009)]{2009ApJ...699.1850B} Berger, E., et al.\ 
2009, \apj, 699, 1850 

\bibitem[Blum et al.(2006)]{2006AJ....132.2034B} Blum, R.~D., et al.\
2006, \aj, 132, 2034

\bibitem[Boersma et al.(2008)]{2008A&A...484..241B} Boersma, C.,
Bouwman, J., Lahuis, F., van Kerckhoven, C., Tielens, A.~G.~G.~M.,
Waters, L.~B.~F.~M., \& Henning, T.\ 2008, \aap, 484, 241

\bibitem[Bonanos et al.(2009)]{2009AJ....138.1003B} Bonanos, A.~Z., et al.\ 
2009, \aj, 138, 1003 

\bibitem[Bond et al.(2003)]{2003Natur.422..405B} Bond, H.~E., et al.\
2003, \nat, 422, 405

\bibitem[Bond et al.(2008)]{2008IAUC.8946....2B} Bond, H.~E., Walter,
F.~M., \& Velasquez, J.\ 2008, \iaucirc, 8946, 2

\bibitem[Bond et al.(2009)]{2009ApJ...695L.154B} Bond, H.~E., Bedin, L.~R., 
Bonanos, A.~Z., Humphreys, R.~M., Monard, L.~A.~G.~B., Prieto, J.~L., 
\& Walter, F.~M.\ 2009, \apjl, 695, L154 

\bibitem[Botticella et al.(2009)]{2009MNRAS.398.1041B} Botticella, M.~T., 
et al.\ 2009, \mnras, 398, 1041

\bibitem[Chandra \& Soderberg(2008)]{2008ATel.1382....1C} Chandra, P.,
\& Soderberg, A.\ 2008, The Astronomer's Telegram, 1382, 1

\bibitem[Davies et al.(2007)]{2007ApJ...671.2059D} Davies, B.,
Oudmaijer, R.~D., \& Sahu, K.~C.\ 2007, \apj, 671, 2059

\bibitem[de Marco(2009)]{2009PASP..121..316D} de Marco, O.\ 2009, \pasp,
121, 316

\bibitem[Dennis et al.(2008)]{2008ApJ...679.1327D} Dennis, T.~J.,
Cunningham, A.~J., Frank, A., Balick, B., Blackman, E.~G., \& Mitran,
S.\ 2008, \apj, 679, 1327

\bibitem[Duley \& Williams(1981)]{1981MNRAS.196..269D} Duley, W.~W., \&
Williams, D.~A.\ 1981, \mnras, 196, 269

\bibitem[Dwek(1983)]{1983ApJ...274..175D} Dwek, E.\ 1983, \apj, 274, 175 

\bibitem[Dwek et al.(1983)]{1983ApJ...274..168D} Dwek, E., et al.\ 1983, 
\apj, 274, 168 

\bibitem[Frost et al.(1998)]{1998ApJ...500..355F} Frost, C.~A.,
Lattanzio, J.~C., \& Wood, P.~R.\ 1998, \apj, 500, 355

\bibitem[Gieren et al.(2005)]{2005ApJ...628..695G} Gieren, W., 
Pietrzy{\'n}ski, G., Soszy{\'n}ski, I., Bresolin, F., Kudritzki, R.-P., 
Minniti, D., \& Storm, J.\ 2005, \apj, 628, 695 

\bibitem[Gogarten et al.(2009)]{2009ApJ...703..300G} Gogarten, S.~M.,
Dalcanton, J.~J., Murphy, J.~W., Williams, B.~F., Gilbert, K., \&
Dolphin, A.\ 2009, \apj, 703, 300

\bibitem[Groenewegen et al.(2007)]{2007MNRAS.376..313G} Groenewegen,
M.~A.~T., et al.\ 2007, \mnras, 376, 313

\bibitem[Hartigan et al.(1987)]{1987ApJ...316..323H} Hartigan, P.,
Raymond, J., \& Hartmann, L.\ 1987, \apj, 316, 323

\bibitem[Houck et al.(2004)]{2004ApJS..154...18H} Houck, J.~R., et al.\
2004, \apjs, 154, 18

\bibitem[Humphreys et al.(2002)]{2002AJ....124.1026H} Humphreys, R.~M., 
Davidson, K., \& Smith, N.\ 2002, \aj, 124, 1026 

\bibitem[Humphreys et al.(2006)]{2006AJ....131.2105H} Humphreys, R.~M.,
et al.\ 2006, \aj, 131, 2105

\bibitem[Ivezic \& Elitzur(1997)]{1997MNRAS.287..799I} Ivezic, Z., \&
Elitzur, M.\ 1997, \mnras, 287, 799

\bibitem[Joblin et al.(2008)]{2008A&A...490..189J} Joblin, C., Szczerba,
R., Bern{\'e}, O., \& Szyszka, C.\ 2008, \aap, 490, 189

\bibitem[Kashi et al.(2009)]{2009arXiv0909.1909K} Kashi, A., Frankowski, 
A., \& Soker, N.\ 2009, arXiv:0909.1909 

\bibitem[Kastner et al.(2006)]{2006ApJ...638L..29K} Kastner, J.~H.,
Buchanan, C.~L., Sargent, B., \& Forrest, W.~J.\ 2006, \apjl, 638, L29

\bibitem[Kastner et al.(2008)]{2008AJ....136.1221K} Kastner, J.~H.,
Thorndike, S.~L., Romanczyk, P.~A., Buchanan, C.~L., Hrivnak, B.~J.,
Sahai, R., \& Egan, M.\ 2008, \aj, 136, 1221

\bibitem[Keller et al.(2008)]{2008ApJ...684..411K} Keller, L.~D., et
al.\ 2008, \apj, 684, 411

\bibitem[Kotak et al.(2005)]{2005ApJ...628L.123K} Kotak, R., Meikle, P.,
van Dyk, S.~D., H{\"o}flich, P.~A., \& Mattila, S.\ 2005, \apjl, 628,
L123

\bibitem[Kotak et al.(2006)]{2006ApJ...651L.117K} Kotak, R., et al.\ 2006, 
\apjl, 651, L117 

\bibitem[Kotak et al.(2009)]{2009ApJ...704..306K} Kotak, R., et al.\ 2009, 
\apj, 704, 306 

\bibitem[Kwok(1993)]{1993ARA&A..31...63K} Kwok, S.\ 1993, \araa, 31, 63 

\bibitem[Kwok et al.(1999)]{1999A&A...350L..35K} Kwok, S., Volk, K., \&
Hrivnak, B.~J.\ 1999, \aap, 350, L35

\bibitem[Kwok et al.(2001)]{2001ApJ...554L..87K} Kwok, S., Volk, K., \&
Bernath, P.\ 2001, \apjl, 554, L87

\bibitem[Marigo(2007)]{2007A&A...467.1139M} Marigo, P.\ 2007, \aap, 467,
1139

\bibitem[Marigo(2008)]{2008MmSAI..79..403M} Marigo, P.\ 2008, Memorie
della Societa Astronomica Italiana, 79, 403

\bibitem[Matsuura et al.(2009)]{2009MNRAS.396..918M} Matsuura, M., et al.\ 
2009, \mnras, 396, 918 

\bibitem[Monard(2008)]{2008IAUC.8946....1M} Monard, L.~A.~G.\ 2008, 
\iaucirc, 8946, 1 

\bibitem[Nomoto(1984)]{1984ApJ...277..791N} Nomoto, K.\ 1984, \apj, 277,
791

\bibitem[Nomoto(1987)]{1987ApJ...322..206N} Nomoto, K.\ 1987, \apj, 322, 
206 

\bibitem[Pastorello et al.(2009)]{2009MNRAS.394.2266P} Pastorello, A.,
et al.\ 2009, \mnras, 394, 2266

\bibitem[Patat et al.(2009)]{2009arXiv0908.0942P} Patat, F., Maund, J.~R., 
Botticella, M.~-., Cappellaro, E., Harutyunyan, A., 
\& Turatto, M.\ 2009, arXiv:0908.0942 

\bibitem[Pathak \& Rastogi(2008)]{2008A&A...485..735P} Pathak, A., \&
Rastogi, S.\ 2008, \aap, 485, 735

\bibitem[Peeters et al.(2002)]{2002A&A...390.1089P} Peeters, E., Hony,
S., Van Kerckhoven, C., Tielens, A.~G.~G.~M., Allamandola, L.~J.,
Hudgins, D.~M., \& Bauschlicher, C.~W.\ 2002, \aap, 390, 1089

\bibitem[Poelarends et al.(2008)]{2008ApJ...675..614P} Poelarends,
A.~J.~T., Herwig, F., Langer, N., \& Heger, A.\ 2008, \apj, 675, 614

\bibitem[Prieto et al.(2008a)]{2008ApJ...681L...9P} Prieto, J.~L., et al.\ 
2008, \apjl, 681, L9 

\bibitem[Prieto(2008b)]{2008ATel.1550....1P} Prieto, J.~L.\ 2008, The 
Astronomer's Telegram, 1550, 1 

\bibitem[Prieto et al.(2008c)]{2008ATel.1596....1P} Prieto, J.~L.,
Kistler, M.~D., Stanek, K.~Z., Thompson, T.~A., Kochanek, C.~S., \&
Beacom, J.~F.\ 2008, The Astronomer's Telegram, 1596, 1

\bibitem[Riera et al.(1995)]{1995A&A...302..137R} Riera, A.,
Garcia-Lario, P., Manchado, A., Pottasch, S.~R., \& Raga, A.~C.\ 1995,
\aap, 302, 137

\bibitem[Riera et al.(2003)]{2003A&A...401.1039R} Riera, A.,
Garc{\'{\i}}a-Lario, P., Manchado, A., Bobrowsky, M., \& Estalella, R.\
2003, \aap, 401, 1039

\bibitem[Roche et al.(1993)]{1993MNRAS.261..522R} Roche, P.~F., Aitken,
D.~K., \& Smith, C.~H.\ 1993, \mnras, 261, 522

\bibitem[Sahai et al.(1999)]{1999ApJ...514L.115S} Sahai, R., Te Lintel
Hekkert, P., Morris, M., Zijlstra, A., \& Likkel, L.\ 1999, \apjl, 514,
L115

\bibitem[Sahai(2009)]{2009astro2010S.256S} Sahai, R.\ 2009, Astronomy,
2010, 256

\bibitem[S{\'a}nchez Contreras et al.(2008)]{2008ApJS..179..166S}
S{\'a}nchez Contreras, C., Sahai, R., Gil de Paz, A., \& Goodrich, R.\
2008, \apjs, 179, 166

\bibitem[Schlegel et al.(1998)]{1998ApJ...500..525S} Schlegel, D.~J., 
Finkbeiner, D.~P., \& Davis, M.\ 1998, \apj, 500, 525 

\bibitem[Siess(2007)]{2007A&A...476..893S} Siess, L.\ 2007, \aap, 476,
893

\bibitem[Sloan et al.(2003)]{2003ApJS..147..379S} Sloan, G.~C., Kraemer,
K.~E., Price, S.~D., \& Shipman, R.~F.\ 2003, \apjs, 147, 379

\bibitem[Sloan et al.(2007)]{2007ApJ...664.1144S} Sloan, G.~C., et al.\
2007, \apj, 664, 1144

\bibitem[Smith et al.(2009)]{2009ApJ...697L..49S} Smith, N., et al.\ 2009, 
\apjl, 697, L49 

\bibitem[Soker(2008)]{2008NewA...13..491S} Soker, N.\ 2008, New
Astronomy, 13, 491

\bibitem[Srinivasan et al.(2009)]{2009AJ....137.4810S} Srinivasan, S., et 
al.\ 2009, \aj, 137, 4810 

\bibitem[Stanghellini \& Renzini(2000)]{2000ApJ...542..308S}
Stanghellini, L., \& Renzini, A.\ 2000, \apj, 542, 308

\bibitem[Stanishev et al.(2008)]{2008CBET.1235....1S} Stanishev, V.,
Pastorello, A., \& Pursimo, T.\ 2008, Central Bureau Electronic
Telegrams, 1235, 1

\bibitem[Steele et al.(2008)]{steele08} Steele, T.~N., et al. 2008,
CBET, 1275, 1

\bibitem[Su{\'a}rez et al.(2006)]{2006A&A...458..173S} Su{\'a}rez, O.,
Garc{\'{\i}}a-Lario, P., Manchado, A., Manteiga, M., Ulla, A., \&
Pottasch, S.~R.\ 2006, \aap, 458, 173

\bibitem[Thompson et al.(2009)]{2009ApJ...705.1364T} Thompson, T.~A., 
Prieto, J.~L., Stanek, K.~Z., Kistler, M.~D., Beacom, J.~F., 
\& Kochanek, C.~S.\ 2009, \apj, 705, 1364 

\bibitem[Tielens(2008)]{2008ARA&A..46..289T} Tielens, A.~G.~G.~M.\ 2008,
\araa, 46, 289

\bibitem[van Loon et al.(1998)]{1998A&A...329..169V} van Loon, J.~T., et
al.\ 1998, \aap, 329, 169

\bibitem[van Marle et al.(2008)]{2008MNRAS.389.1353V} van Marle, A.~J., 
Owocki, S.~P., \& Shaviv, N.~J.\ 2008, \mnras, 389, 1353 

\bibitem[Vijh et al.(2009)]{2009AJ....137.3139V} Vijh, U.~P., et al.\ 2009, 
\aj, 137, 3139 

\bibitem[Wesson et al.(2008)]{2008CBET.1381....1W} Wesson, R., Fabbri,
J., Barlow, M., \& Meixner, M.\ 2008, Central Bureau Electronic
Telegrams, 1381, 1

\bibitem[Wesson et al.(2009)]{wesson09} Wesson, R., et al.\ 2009,
\mnras, submitted, arXiv:0907.0246

\bibitem[Witt et al.(2009)]{2009ApJ...693.1946W} Witt, A.~N., Vijh,
U.~P., Hobbs, L.~M., Aufdenberg, J.~P., Thorburn, J.~A., \& York, D.~G.\
2009, \apj, 693, 1946

\bibitem[Wood et al.(1992)]{1992ApJ...397..552W} Wood, P.~R., Whiteoak,
J.~B., Hughes, S.~M.~G., Bessell, M.~S., Gardner, F.~F., \& Hyland,
A.~R.\ 1992, \apj, 397, 552

\bibitem[Wooden et al.(1993)]{1993ApJS...88..477W} Wooden, D.~H., Rank,
D.~M., Bregman, J.~D., Witteborn, F.~C., Tielens, A.~G.~G.~M., Cohen,
M., Pinto, P.~A., \& Axelrod, T.~S.\ 1993, \apjs, 88, 477

\bibitem[Yee et al.(2008)]{2008CBET.1340....1Y} Yee, J.~C., Eastman,
J.~D., \& Prieto, J.~L.\ 2008, Central Bureau Electronic Telegrams,
1340, 1

\end{thebibliography}
\end{document}